\documentclass[10pt, twocolumn, pre, aps, superscriptaddress, showpacs]{revtex4-1}
\usepackage{amsmath, graphicx, subfigure, tikz}
\begin{document}

    \title{A Lower Lower-Critical Spin-Glass Dimension from\\
    Quenched Mixed-Spatial-Dimensional Spin Glasses}
    \author{Bora Atalay}
    \affiliation{Faculty of Engineering and Natural Sciences, Sabanc\i\ University, Tuzla, Istanbul 34956, Turkey}
    \author{A. Nihat Berker}
    \affiliation{Faculty of Engineering and Natural Sciences, Kadir Has University, Cibali, Istanbul 34083, Turkey}
    \affiliation{Department of Physics, Massachusetts Institute of Technology, Cambridge, Massachusetts 02139, USA}



\begin{abstract}

By quenched-randomly mixing local units of different spatial
dimensionalities, we have studied Ising spin-glass systems on
hierarchical lattices continuously in dimensionalities $1 \leq d
\leq3$. The global phase diagram in temperature, antiferromagnetic
bond concentration, and spatial dimensionality is calculated. We
find that, as dimension is lowered, the spin-glass phase disappears
to zero temperature at the lower-critical dimension $d_c=2.431$. Our
system being a physically realizable system, this sets an upper
limit to the lower-critical dimension in general for the Ising
spin-glass phase. As dimension is lowered towards $d_c$, the
spin-glass critical temperature continuously goes to zero, but the
spin-glass chaos fully sustains to the brink of the disappearance of
the spin-glass phase. The Lyapunov exponent, measuring the strength
of chaos, is thus largely unaffected by the approach to $d_c$ and
shows a discontinuity to zero at $d_c$.

\end{abstract}
\maketitle

\section{Introduction: Spin-Glass Lower-Critical Dimension}

The lower-critical dimension $d_c$ of an ordering system, where the
onset of an ordered phase is seen as spatial dimension $d$ is
raised, has been of interest as a singularity of a continuous
sequence of singularities, the latter being the phase transitions to
the ordered phase which change continuously as $d$ is raised from
$d_c$.  The lower-critical dimension of systems without quenched
randomness has been known for some time as $d_c=1$ for the
Ising-type ($n=1$ component order-parameter) systems, $d_c=2$ for
XY, Heisenberg, ... ($n=2,3,..$) systems, highlighted with a
temperature range of criticality at $d_c=2$ of the XY model
\cite{StanleyKaplan,Stanley}. In systems with quenched randomness, a
marvelous controversy on the lower-critical dimension of the
random-field Ising system has settled for
$d_c=2$.\cite{Jaccarino,Birgeneau,Wong,Berker84,Aizenman,AizenmanE,Machta,
Falicov} Quenched bond randomness affects the first- versus
second-order nature of the phase transition into an ordered phase
that exists without quenched randomness (such as the ferromagnetic
phase), rather than the dimensional onset of this ordered phase.

\begin{figure}[ht!]
\centering
\includegraphics[scale=0.25]{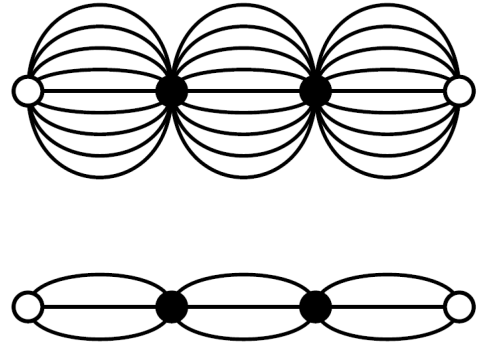}
\caption{Local graphs with $d = 2$ (bottom) and $d = 3$ (top)
connectivity.  The cross-dimensional hierarchical lattice is
obtained by repeatedly imbedding the graphs in place of bonds,
randomly with probability $1-q$ and $q$ for the $d = 2$ and $d = 3$
units, respectively.}
\end{figure}

\begin{figure*}[ht!]
\centering
\includegraphics[scale=0.8]{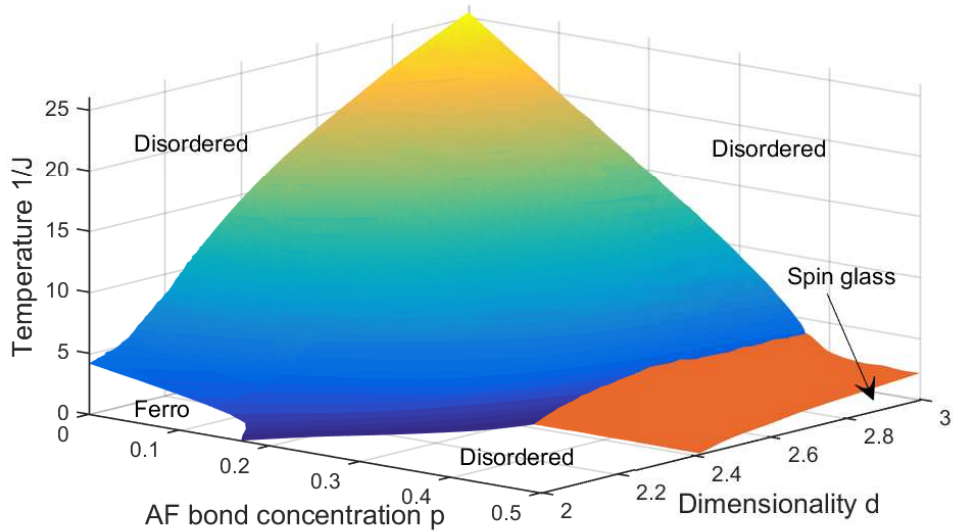}
\caption{Calculated exact global phase diagram of the Ising spin
glass on the cross-dimensional hierarchical lattice, in temperature
$1/J$, antiferromagnetic bond concentration $p$, and spatial
dimension $d$. The global phase diagram being symmetric about $p =
0.5$, the mirror image portion of $0.5 < p < 1$ is not shown.  The
spin-glass phase is thus clearly seen, taking off from zero
temperature at $d_c = 2.431$.}
\end{figure*}

The situation is inherently different with an ordered phase that is
caused by the quenched randomness of competing
ferromagnetic-antiferromagnetic (and more recently right-left
chirality or helicity \cite{Caglar3}) interactions, namely the Ising
spin-glass phase.  Replica-symmetry-breaking mean-field theory
yields $d_c=2.5$,\cite{Parisi} this being of immediate high interest
as the first known example of a non-integer lower-critical
dimension. Numerical fit to spin-glass critical temperatures
\cite{Boettcher} and free energy barriers \cite{Parisi2} for integer
dimensions also suggests $d_c=2.5$. Numerical fits to the exact
renormalization-group solutions of two different families of
hierarchical lattices with a sequence of decreasing dimensions yield
$d_c= 2.504$ (Ref.\cite{Amoruso, Bouchaud}) and $d_c= 2.520$
(Ref.\cite{Demirtas}), which are of further interest by being
non-simple fractions. The strength of hierarchical lattice
approaches is that they present exact (numerical) solutions
\cite{BerkerOstlund,Kaufman1,Kaufman2}, but they involve non-unique
continuations between integer dimensions, being based on different
families of fractal graphs. However, in the hunt for the
lower-critical dimension, since each hierarchical lattice
constitutes a physical realization, calculating a finite-temperature
spin-glass phase at $d$ automatically pushes the lower-critical
dimension to $d_c<d$, which is an important piece of information.

The exact numerical renormalization-group solution of hierarchical
lattices, used in the current study, has been fully successful in
all aspects of lower-critical-dimension behavior mentioned in the
first paragraphs of this Section.  Whereas previous studies with
hierarchical lattices have used in each calculation a lattice with
the same dimensionality at every locality (these include but are not
confined to hierarchical lattices that are simultaneously
approximate solutions \cite{Migdal,Kadanoff} for hypercubic and
other Euclidian lattices), we quenched randomly mix units with local
dimensionality $d=2$ and $d=3$.  By varying the relative
concentration of these two units, we continuously span from $d=2$ to
$d=3$.  In this physically realized system, we find $d_c=2.431$,
lower than previously found values and thus setting an upper limit
to the actual lower-critical dimension of the Ising spin-glass
phase.  Furthermore, as our spin-glass phase disappears at
zero-temperature at $d_c=2.431$, it is fully chaotic, with a
calculated Lyapunov exponent of $\lambda=1.56$ (this exponent equals
1.93 at $d=3$), which is in sharp contrast to the disappearance, as
frustration is microscopically turned off, of the spin-glass phase
to the Mattis-gauge-transformed ferromagnetic phase, where the
Lyapunov exponent (and chaos) continuously goes to
zero.\cite{Ilker2} In the current work, we also obtain a global
phase diagram in the variables of temperature, antiferromagnetic
bond concentration, and spatial dimensionality.

\section{Model and Method: Moving between Spatial Dimensions through Local Differentiation}

The Ising spin-glass system has Hamiltonian
\begin{equation}
-\beta \mathcal{H}=\sum_{\langle ij \rangle} J_{ij} s_i s_j
\end{equation}
where $\beta=1/kT$, at each site $i$ of the lattice the spin $s_i =
\pm 1$, and $\langle ij \rangle$ denotes summation over all
nearest-neighbor site pairs. The bond $J_{ij}$ is ferromagnetic
$+J>0$ or antiferromagnetic $-J$ with respective probabilities $1-p$
and $p$.  This Hamiltonian is lodged on the hierarchical lattice
constructed with the two graphs shown in Fig. 1.  The lower graph
has a length rescaling factor (distance between the external
vertices) of $b=3$ and a volume rescaling factor (number of internal
bonds) of $b^d = 9$.  Thus, self-imbedding the lower graph into its
bonds \textit{ad infinitum} results in a $d=2$ spatial dimensional
lattice that is numerically exactly soluble.  The upper graph
similarly yields $d=3$.  Other graphs have been used to
systematically obtain intermediate non-integer dimensions
\cite{Demirtas}.

\begin{figure}[ht!]
\centering
\includegraphics[scale=0.9]{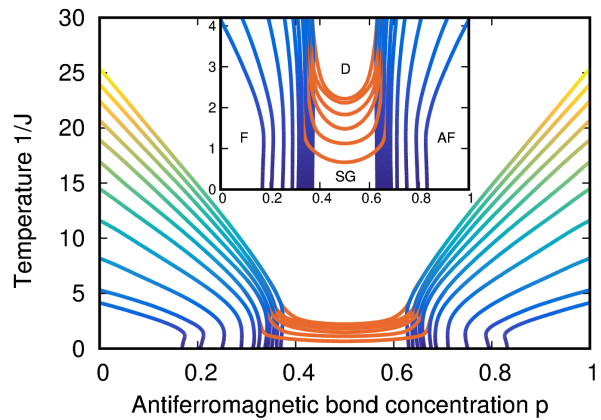}
\caption{Constant dimensionality $d$ cross sections of the global
phase diagram in Fig. 2.  The cross sections are, starting from high
temperature, for $d=3, 2.9, 2.8, 2.7, ..., 2.1, 2$. It is seen that,
as the dimensionality $d$ approaches $d_c =2.431$ from above, the
spin-glass phase disappears at zero temperature.}
\end{figure}

\begin{figure}[ht!]
\centering
\includegraphics[scale=0.9]{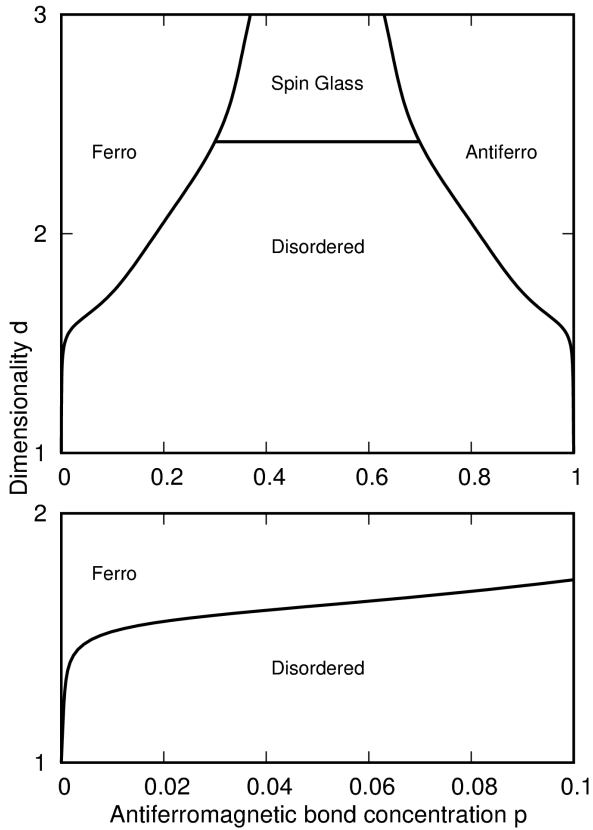}
\caption{Zero-temperature phase diagram of the Ising spin-glass
system on the cross-dimensional hierarchical lattice, in
antiferromagnetic bond concentration $p$ and and spatial dimension
$d$.  The lower-critical dimension of $d_c = 2.431$ is clearly
visible.}
\end{figure}

For recent exact calculations on hierarchical lattices, see
Refs.\cite{Masuda, Boettcher4, Bleher, Zhang2017, Peng, Sirca, Maji,
Boettcher5, Atalay}. Thus, previous works have generally used a
hierarchical lattice generated by a single graph and spatial
dimensionality that is microscopically uniform throughout the
system.  By contrast, we mix the two graphs with local $d=2$ and
$d=3$ in frozen randomness and definite proportionality:  Starting
with either graph (in the thermodynamic limit, this choice does not
matter), each bond is replaced by the $d=2$ or $d=3$ graph, with
probability $1-q$ and $q$, respectively.  This random imbedding is
repeated \textit{ad infinitum}. Thus, the dimensionality of the
macroscopic system is $(1-q)\times 2 + q\times 3 = 2+q$.

The exact renormalization-group solution of this system works in the
opposite direction from the lattice construction just described.  As
described after Eq.(1), we start with the double-valued distribution
of $+J$ or $-J$ bonds, with probabilities $1-p \,$  and $p \,$
respectively, on a $d=2$ or $d=3$ unit with probabilities $1-q$ and
$q$ respectively. The local renormalization-group transformation
proceeds by $b^{d-1}$ bond-movings followed $b=3$ (to preserve the
ferromagnetic-antiferromagnetic symmetry) decimations, generating a
distribution of 500 new interactions, which is of course no longer
double valued.\cite{Atalay}  (In fact, for numerical efficiency,
these operations are broken down to binary steps, each involving two
distributions of 500 interactions.)  In the disordered phase, the
interactions converge to zero.  In the ferromagnetic and
antiferromagnetic phases, under renormalization-group, the
interaction diverges to strong coupling as the renormalized average
$\overline{J}\, ' \sim b^{y_R^F} \overline{J}$, where the prime
refers to the renormalized system and $y_R^F>0$ is the runaway
exponent of the ferromagnetic sink of the renormalization-group
flows.  In the spin-glass phase, under renormalization-group, the
distribution of interactions continuously broadens symmetrically in
ferromagnetism and antiferromagnetism, the absolute value of the
interactions diverging to strong coupling as the renormalized
average $\overline{|J|}\, ' \sim b^{y_R^{SG}} \overline{|J|}$, where
$y_R^{SG}>0$ is the runaway exponent of the spin-glass sink of the
renormalization-group flows. The runaway exponents $y_R^F$ and
$y_R^{SG}$ are given below as a function of dimensionality $d$.

\begin{figure}[ht!]
\centering
\includegraphics[scale=0.9]{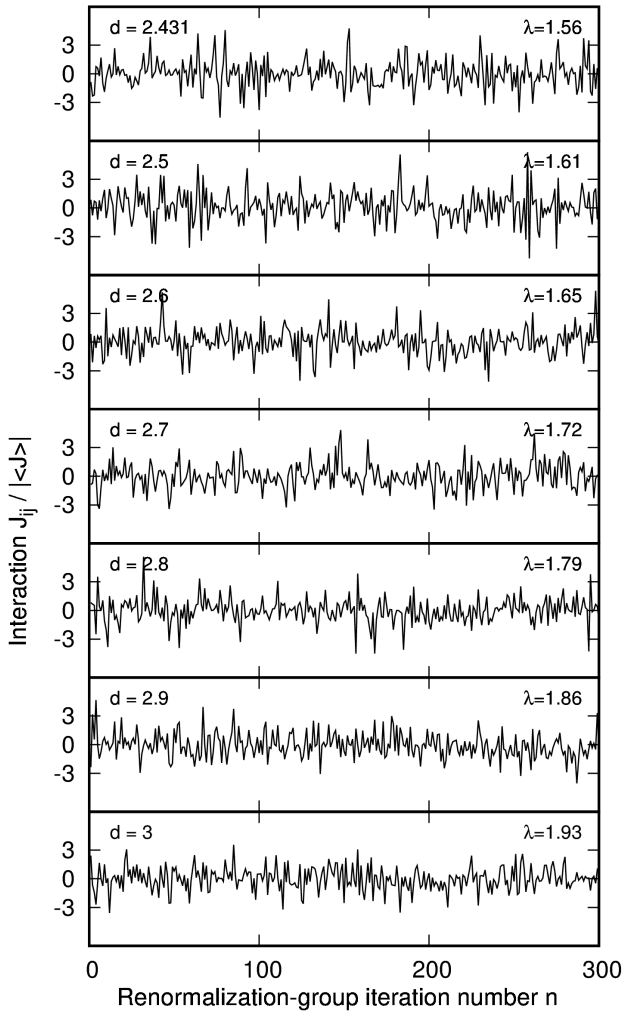}
\caption{The chaotic renormalization-group trajectory of the
interaction $J_{ij}$ at a given location $<ij>$, for various spatial
dimensions between the lower-critical $d_c = 2.431$ and $d = 3$.
Note that strong chaotic behavior, as also reflected by the shown
calculated Lyapunov exponents $\lambda$, nevertheless continues as
the spin-glass phase disappears at the lower-critical dimension
$d_c$, as also seen in Fig. 6.}
\end{figure}

\begin{figure}[ht!]
\centering
\includegraphics[scale=0.9]{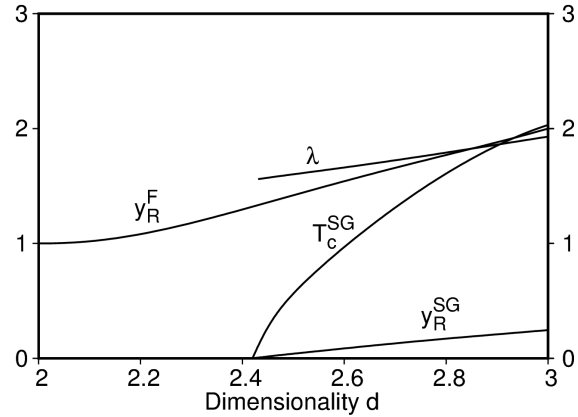}
\caption{Spin-glass critical temperature $T_C^{SG}$ at $p = 0.5$,
spin-glass chaos Lyapunov exponent $\lambda$, spin-glass-phase
runaway exponent $y_R^{SG}$ and ferromagnetic-phase runaway exponent
$y_R^{F}$, as a function of dimension $d$.  Note that the
ferromagnetic phase runaway exponent $y_R^{F}$ correctly tracks
$d-1$.}
\end{figure}

\section{Transitional Dimensional Global Phase Diagram and Full Chaos Even at Spin-Glass Disappearance}

Figure 2 shows our calculated global phase diagram in the variables
of temperature $1/J$, antiferromagnetic bond concentration $p$, and
spatial dimensionality $2 \leq d \leq3$.  In addition to the
high-temperature disordered phase, ferromagnetic, antiferromagnetic
(the phase diagram being ferromagnetic-antiferromagnetic symmetric
about $p=0.5$, the mirror-image antiferromagnetic part of $p>0.5$ is
not shown; however, see Figs. 3 and 4), and spin-glass ordered
phases are seen. As dimensionality $d$ is lowered, the spin-glass
phase disappears at zero temperature at the lower-critical dimension
of $d_c =2.431$. Constant-dimension $d$ cross sections of the global
phase diagram are in Fig. 3, where the gradual temperature-lowering
of the spin-glass phase, as the lower-critical dimension $d_c =
2.431$ is approached from above, is seen.  However, such gradual
disappearance is not the case for the chaos \cite{McKayChaos,
McKayChaos2,BerkerMcKay} inherent to the spin-glass phase, as seen
below.

Fig. 4 shows the calculated zero-temperature phase diagram in the
variables of antiferromagnetic bond concentration $p$ and spatial
dimensionality $1 \leq d \leq3$.  For this Figure, the calculation
is continuously extended down to $d = 1$ by again quenched-randomly
mixing our $d=2$ graph (Fig. 1) and a linear 3-segment strand.  The
smoothness of the boundaries at $d=2$ validates our method.  The
independence of $d_c$ from $p$ is noteworthy.

An inherent signature of the spin-glass phase is the chaotic
behavior \cite{McKayChaos, McKayChaos2,BerkerMcKay,
ZZhu,Katzgraber3,Fernandez6,Fernandez8, Wang2} of the interaction at
a given locality as a function of scale change, namely under
consecutive renormalization-group transformations. This chaos is
shown in Fig. 5 for a variety of dimensions, including the
lower-critical dimension $d_c=2.431$.  For each chaos, the Lyapunov
exponent
\begin{equation}
\lambda = \lim _{n\rightarrow\infty} \frac{1}{n} \sum_{k=0}^{n-1}
\ln \Big|\frac {dx_{k+1}}{dx_k}\Big|
\end{equation}
where $x_k = J(ij)/\overline{|J|}$ at step $k$ of the
renormalization-group trajectory, measures the strength of the
chaos, and is calculated and shown for the spatial dimensions in
Fig. 5.  It is seen that the system shows strong chaos (positive
Lyapunov exponent $\lambda=1.56$) even at $d_c=2.431$, namely at the
brink of the disappearance of the spin-glass phase, after an
essentially slow numerical evolution from the $d=3$ value of
$\lambda=1.93$. This is in sharp contrast with the disappearance of
the spin-glass phase, into a Mattis-gauge-transformed ferromagnetic
phase, as frustration is gradually turned off microscopically, where
chaos gradually disappears and the Lyapunov exponent continuously
goes to zero, as seen in Fig. 6 of Ref. \cite{Ilker2}. As seen in
Fig. 6, the Lyapunov exponent, shown continuously as a function of
dimension, is essentially unaffected by the disappearance of the
spin-glass phase and thus shows a discontinuity at $d_c$.  The
runaway exponent of the spin-glass phase, on the other hand,
correctly goes to zero at $d_c$, as is expected by the
renormalization-group flow structure. Also seen in Fig. 6 is the
spin-glass critical temperature going to zero at $d_c$.

\section{Conclusion: Lower Lower-Critical Dimension and Lyapunov Discontinuity}

By quenched-randomly mixing local units of different spatial
dimensionalities, we have studied Ising spin-glass systems on
hierarchical lattices continuously in dimensionalities $1 \leq d
\leq3$. We have calculated the global phase diagram in temperature,
antiferromagnetic bond concentration, and spatial dimensionality. We
find that, as dimension is lowered, the spin-glass phase disappears
at zero temperature at $d_c=2.431$.  Our system being a physically
realizable system, this sets an upper limit to the lower-critical
dimension of the Ising spin-glass phase.  As dimension is lowered
towards $d_c$, the spin-glass critical temperature continuously goes
to zero.  The Lyapunov exponent, measuring the strength of chaos, is
on the other hand largely unaffected by the approach to $d_c$ and
shows a discontinuity to zero at $d_c$.

\begin{acknowledgments}
Support by the Academy of Sciences of Turkey (T\"UBA) is gratefully
acknowledged. We thank Tolga \c{C}a\u{g}lar for most useful
discussions.
\end{acknowledgments}

\end{document}